\documentclass[12pt]{article}
\begin{document}
\begin{center}
{\bf Is torsion needed in a theory of gravity? A reappraisal}
\end{center}
\vspace{.2cm}
\begin{center}
Janusz Garecki\footnote{e-mail address:
garecki@wmf.univ.szczecin.pl}
\end{center}
\vspace{.2cm}
\begin{center}
Institute of Mathematics, University of Szczecin, Wielkopolska 15,
70--451 Szczecin, POLAND
\end{center}
\date{\today}
\vspace{.3cm}
\begin{center}
This paper is written Lecture delivered at {\bf
Hypercomplex Seminar 2011} in B\c edlewo (Poland), 23-30 July 2011.
The Lecture was an extended and updated version our old lectures and an article published in
2004.\cite{Gar04}\\
{\bf In Memory of Professor Roman S. Ingarden who was died July 12,
2011.}
\end{center}
\begin{abstract}
It is known that General Relativity ({\bf GR}) uses a Lorentzian
Manifold $(M_4;g)$ as a geometrical model of the physical
spacetime. The metric $g$ is required to satisfy
Einstein's  equations. Since the 1960s many authors have tried to generalize this
geometrical model of the physical space--time by introducing torsion.
In this paper we discuss the present status of torsion in a theory of gravity.
Our conclusion is that the general--relativistic model of the physical
spacetime is sufficient for the all physical applications and it seems to be the most
satisfactory.
\end{abstract}
\newpage
\section{Introduction}

In past we were enthusiast of torsion, mainly under influence of
excellent papers given by F.W. Hehl and A. Trautman. But studying
{\it Poincare' field theories of gravity} ({\bf PGT}) one can
easily see that torsion leads to serious complications,
especially calculational.

About twenty years ago we have observed that the our idea of the
{\it superenergy and supermomentum tensors} (very effective in general relativity)
fails in a spacetime having torsion. So, our interest to torsion has diminished.

In the meantime we have read many papers by C.M. Will, G. Esposito-Farese, T.
Damour, S. Kopeikin, S.G. Thuryshev and others devoted recent
experiments on gravity. As we understood all these experiments
confirmed standard general relativity ({\bf GR}) with a very high
precision and excluded torsion, at least in vacuum.

Besides, during the last three decades there was given many interesting
papers on {\it universality} of the {\bf GR}  equations. So, in
consequence, we have decided to analyze  status of torsion  in
gravitational physics. From this analysis the review has originated.

Of course, {\it we do not prove} that torsion  is not admissible at all.
Rather, we only give short information about recent gravitational experiments and
collect problems which arise when one introduces torsion as a part of the geometrical
structure of the physical spacetime.
But, as you see, we will finish review with the conclusion (based mainly
on {\it Ockham razor}):
\begin{enumerate}

\item Torsion in needn't in a theory of gravity:
\item The Levi-Civita connection is sufficient for the all
physical applications. This the most simple connection is exactly
just what we need.
\end{enumerate}

The paper is organized as follows. In Section II we remind a
general definition of torsion and in Section III we consider
motivations to introduce torsion into geometrical model of the
physical spacetime. We will see that these motivations {\it are
not convincing}. In Section IV we very shortly discuss
experimental evidence for torsion and Section V we present
arguments against torsion in a theory of gravity. We will conclude
in Section VI (from the facts given in the two previous Sections) that torsion rather {\it
should not be introduced into a geometrical model of the physical
spacetime}.

\section{Torsion of a linear connection $\omega^i_{~k}$ on $L(M)$}
\vspace{.3cm}

We confine to the metric-compatible connection which satisfies
$Dg_{ik} = dg_{ik} -\omega^p_{~i}g_{pk} - \omega^p_{~k} g_{ip}=
0$ because we do not see any reasons to consider more general connection.
Here, and in the following, $D$ means exterior covariant derivative and $d$ is the ordinary
exterior derivative.

One can give the following, general definition of torsion $\Theta^i$
\cite{Kob63,Trau84} of a linear connection
\begin{equation}
\Theta^i :=D\theta^i = d\theta^i + \omega^i_{~k}\wedge\theta^k =:
{1\over 2}Q^i_{~kl}\theta^k\wedge\theta^l.
\end{equation}
Here $\theta^i$ are {\it canonical 1-forms} (or {\it soldering
1-forms}) on the principial bundle of the linear frames $L[M,GL(n;R),\pi]$
($L(M)$ in short) over a manifold $M$, and $Q^i_{~kl}$ denote
components of the torsion tensor.

After pulling back by local section $\sigma:U\rightarrow L(M);~~U\subset M$, one
gets on the base M
\begin{equation}
{\tilde\Theta}^i = d\vartheta^i
+{\tilde\omega}^i_{~k}\wedge\vartheta^k = {1\over 2}{\tilde
Q}^i_{~kl}\vartheta^k\wedge\vartheta^l.
\end{equation}
${\tilde\omega}^i_{~k}:= \sigma_{\ast}\omega^i_{~k}$
are pull-back of $\omega^i_{~k}$ and  $\vartheta^i$ are pull-back of
$\theta^i$. $\vartheta^l := \vartheta^0,~\vartheta^1,~\vartheta^2,~\vartheta^3$
form a Lorentzian coframe on $M$.

In a coordinate (= holonomic) frame $\{\partial_i\}$ and dual coframe
$\{dx^k\}$ on $M$ one has ${\tilde\omega}^i_{~k} =
\Gamma^i_{~lk}dx^l$, and ${\tilde Q}^i_{~kl} = \Gamma^i_{~kl} -
\Gamma^i_{~lk}$.
\section{Motivation to introduce of torsion into gravity}
In the 1960s--1970s  some researchers introduced torsion into the theory of gravity\footnote{We omit
here older attempts to introduce torsion because they have only historical
meaning.}\cite{Ut56,Kib61,Sci62,Tra73,Heh73,Heh76}.
The main motives (only theoretical) were  the following:
\begin{enumerate}
\item Studies on geometric theory of dislocations (Theory
of a generalized Cosserat continuum) led, following G\"unter, Hehl, Kondo, and
Kr\"oner, to heuristic arguments for a metric spacetime with torsion, i.e.,
to Riemann--Cartan spacetime.
\item Investigations of spinning matter in {\bf GR} resulted in conclusion
that the canonical energy--momentum tensor of matter $_c T_i^{~ k}$ can be source
of curvature and the canonical intrinsic spin density tensor $_c S^{ikl} = (-)
_c S^{kil}$ can be source of torsion of the underlying spacetime. From
this Einstein--Cartan--Sciama--Kibble ({\bf ECSK}) theory and its
generalizations originated.
\item Attemts to formulate gravity as a gauge theory for Lorentz group
$L$ or for Poincare' group $P$ by using {\it Palatini's formalism} led to a space--time
endowed with a metric--compatible connection which might have (but not necessarily)
non--vanishing torsion, i.e., again one was led to Riemann--Cartan space--time
\cite{Heh80,Tra80,Hay80,Gar90,Tra99}.
\end{enumerate}

Some remarks are in order concernig 3.
\begin{enumerate}
\item If we admit a metric -compatible connection with torsion
when ``gauging'' groups $L$ or $P$ by using Palatini's approach and Ehreshmann
theory of connection, then we will end up with
strange situation, different then in ordinary gauge fields: we get
a ``gauge theory'' which has two gauge potentials
\begin{equation}
\vartheta^i~---translational~ (=pseudoorthonormal~ coframe),
\end{equation}
\begin{equation}
\omega^i_{~k} ~---rotational ~(= metric-compatible~ linear
~connection),
\end{equation}
and two gauge strengths
\begin{equation}
\Theta^i= D\vartheta^i~---translational~ (torsion),
\end{equation}
\begin{equation}
\Omega^i_{`k} = D\omega^i_{~k}~--- rotational~ (curvature).
\end{equation}
Notice that $\vartheta^i$ {\it do not transform} like gauge
potentials and contribute to
$\omega^i_{~k} = {_{LC}\omega}^i_{~k} + K^i_{~k}$; besides,
the {\it gauge strengths}
$\Theta^i = K^i_{~k}\wedge\vartheta^k$
{\it contribute} to $\omega^i_{~k} = {_{LC}\omega}^i_{~k} +
K^i_{~k}$ and also to $\Omega^i_{~k}   = d\omega^i_{~k}
+\omega^i_{~p}\wedge\Omega^p_{~k}$.

Here $_{LC}\omega^i_{~k}$ denotes the {\it Levi-Civita Connection}
and $K^i_{~k}$ is the contortion.

So, the gauge potentials and gauge strengths {\it are not
independent} in the case. This is not satisfactory and suggests {\it other approach to
``gauging'' gravity}.
\item Besides, the action integrals in these trials to  gauge gravity {\it
didn't have} forms like an action integral for a gauge field, $\int
tr(F\wedge\star F)$, and led to very complicated field equations of
3rd order, different from {\bf GR} equations. These field equations
contain many arbitrary parameters ( 10 apart from
$\Lambda$ in the case of the so-called Poincare'  Gravity
Theories, {\bf PGT}). Here $\star$ means Hodge duality operator.

There exist many serious problems connected with these field
equations: tachyons, ghosts, instability of their solutions,
ill--posedness Cauchy problem, etc., (see, e.g.,\cite{Str}).

We would like to emphasize that there exists an old approach to
``gauge'' gravity proposed by Yang \cite{Yang74} which has action
 typical for a gauge field: $\int
 \Omega^i_{~k}\wedge\star\Omega^k_{~i}$. But, unfortunately, this
 approach leads to incorrect theory of gravity.
\end{enumerate}

The above theoretical motives  {\it are not convincing}. For example, the often used argument
for torsion (following from study of spinning matter in {\bf GR}) based on the (non-homogeneous)
holonomy theorem \cite{Tra73,Tra99}\footnote{This theorem says that torsion gives
translations, and curvature gives Lorentz rotations in tangent spaces of
a Riemann-Cartan manifold induced by (Cartan) displacement along loops.} holds only if
one uses {\it Cartan displacement} which displaces  vectors and contactpoints \cite{Sch54}.
Ordinary parallell displacement (which displaces only vectors) gives only Lorentz rotations
(= homogeneous holonomy group) even in a Riemann-Cartan space-time \cite{Sch54}.
Moreover, there are other geometrical interpretations of torsion, e.g., Bompiani
\cite{Bom51} {\it connects torsion with rotations in tangent spaces}, not with translations.

We also needn't to generalize {\bf GR} in order to get a gauge theory with $L$ or
$P$ as a gauge group \cite{Gren83,Sab85,Wiese010}. The most convincing argument in this field
is given by {\it Cartan's approach} to connection and geometry
\cite{Wiese010}.

Roughly speaking, in Cartan's approach (for details, see \cite{Wiese010}) one combines
the linear Ehreshmann connection form $\omega$ and coframe field $\theta$ into one
connection $A = \omega\oplus\theta$ valued in a larger Lie
algebra ${\bf g}$ (In our case $\omega$ is the Ehresmann connection
on principal bundle of the orthonormal frames $O[M,L,\pi]$ and $\theta$ is the soldering
form on this bundle. ${\bf g}$ is the algebra of the Poincare'
group $P$ or de Sitter group).

In consequence, one has only one gauge potential $A =\omega\oplus\theta$
and one gauge strength ${\hat F} = \Omega - {\Lambda\over 3}\theta\wedge\theta$
($\Lambda$ is the cosmological constant) for gravity.

Using Cartan's approach to connection one can write the ordinary
{\bf GR} action with $\Lambda$, $S_g = \int \sqrt{\vert g\vert}\bigl(R -\Lambda)
d^4x$, in the form $S_g = (-){3\over 2G\Lambda}\int tr({\hat F}\wedge\star{\hat
F})$, i.e., {\it exactly} in the form of the action of a gauge
field.

Thus, the Cartan's  (not Ehreshmann) approach to connection and geometry suits to
correct ``gauging'' of {\bf GR}. The Ehreshmann theory suits to
ordinary gauge fields.

There exists also an other approach to {\bf GR} as a gauge theory developed
by A. Ashtekar, C. Rovelli, J. Lewandowski and  covorkers (Ashtekar's variables)
\cite{Giu94,Ash99,Iri88,Rov20}. In this approach {\bf GR} is also very akin
to a Yang--Mills theory.

In resuming, one can say that {\it we needn't generalize or modify} {\bf
GR} in order to obtain a gauge theory of gravity.
\section{Experimental evidence for torsion}

Up to now we {\it have no experimental evidence} for existence
torsion in Nature (see, e.g., \cite{Shap}). There exist only very
{\it stringent constraints} on torsion components obtained in a
speculative, purely theoretical, methods (see, eg., \cite{Shap, Ricc11}).

 To the contrary, all gravitational experiments confirmed with a very
high precision ($ \sim 10^{-14}$) Einstein's Equivalence Principle ({\bf
EEP}) and, with a smaller precision (up to $0,003\%$ in Solar System , i.e., in weak field,
and up to $0,05\%$ in binary pulsars, i.e., in strong
gravitational fields) the General Relativity ({\bf GR}) equations
\cite{Clif93,Esp99,Dam99,Slava,Kop,Kram,Gad}.
Here by {\bf EEP} we mean a formulation of this Principle given by C. W.
Will \cite{Clif93}. In this formulation\footnote{This constructive formulation of the
Principle can be experimentally tested.} the  {\bf EEP} states:
\begin{enumerate}
\item The Weak Equivalence Principle ({\bf WEP}) is valid.
This means that the trajectory of a freely falling spherical test body (one
not acted upon by non-gravitational forces and being too small to be affected
by tidal forces) {\it is independent of its internal structure and
material composition}.
\item Local Lorentz Invariance ({\bf LLI}) is valid.
This means that the outcome of any local non-gravitational experiment
{\it is independent of the velocity of the freely-falling and non-rotating reference frame in
which it is performed}.
\item Local Position Invariance ({\bf LPI}) is valid.
This means that the outcome of any local non-gravitational experiment
performed in a freely-falling and non-rotating reference frame
{\it is independent of where and when in the Universe it is performed}.
\end{enumerate}

Following C.M. Will, the only theories of gravity that can embody {\bf EEP} in the above
constructive formulation of the Principle are those that
satisfy the postulates of {\it metric theories of gravity} \cite{Clif93}, which
are:
\begin{enumerate}
\item Spacetime is endowed with a symmetric metric.
\item The trajectories of freely falling spherical test bodies are geodesics of
that metric.
\item In local freely-falling and non-rotating reference frames the non-gravitational
laws of physics are those written in the language of Special Relativity
({\bf SRT}).
\end{enumerate}

C.M. Will called the  {\bf EEP} ``heart and soul of {\bf GR}''.

The {\bf EEP} implies a  {\it universal pure metric coupling between
matter and gravity}. It  admits {\bf GR}, of course, and, at most, some of
the so--called {\it scalar--tensor theories} (these, which respect {\bf EEP})
\cite{Clif93,Esp99,Dam99} without torsion.

So, torsion seems to be  {\it excluded} in vacuum or at least {\it very strongly constrained}
in vacuum by the latest gravitational experiments , i.e., at least propagating torsion
is excluded or very strongly constrained by these experiments \footnote{Torsion is excluded
or very strongly constrained at least in vacuum because if we neglect a cosmological background,
then the all gravitational experiments were performed in vacuum. This means that {\bf ECSK}
theory can survive since this theory is identical in vacuum with {\bf GR}. Of
course, the same is true for other gravity theories which in vacuum
reduce to {\bf GR}. But the gravity theories of such a kind do not admit propagating free
torsion.} which have confirmed {\bf EEP} with very high precision.
As a consequence, at least freely propagating torsion still seems to be
{\it purely hypothetical}.

We would like to emphasize that T. Damour already concluded in past \cite{Dam99}:
``Einstein was right at least $99.9999999999\%$ concerning {\bf EEP} and $99,9\%$
concerning Lagrangian and field equations''.

Thus, from the experimental point of view, up to now, torsion {\it
is needn't} in a theory of gravity.
\section{Theoretical arguments against torsion}

We begin this Section with the remark that if one utilizes the
so--called ``Ockham's razor''\footnote{By ``Ockham's razor'' we mean a
Philosophical Principle which states: ``Entities are not to be
multiplied without necessity''.} then torsion is needn't for him in a
theory of gravity because the {\it wonderful, the most simple and most
symmetric} Levi--Civita connection is sufficient for the all physical
requirements.

The first our argument against torsion is given in the very important paper
by J. Ehlers, F.A.E. Pirani , and A. Schild \cite{EPS}. These
authors have showed that requiring compatibility between conformal
 geometry ${\bf C}$ defined by rays of light and the projective
 structure ${\bf P}$ of spacetime determined by trajectories  of
 freely-falling test particles leads to Weyl spacetime with a
{\it symmetric connection} $\omega$. Then, admitting some, very
natural axioms \cite{EPS}, we obtain Riemannian geometry.

So, studying the rays of light and freely--falling particles,
leads us to Riemannian spacetime.

Now, let us pay our attention to the other, disadvantegeous properties of
torsion and metric--compatible spacetimes with torsion:
\begin{enumerate}
\item In a spacetime with torsion {\it do not exist} infinitesimal
parallelograms \cite{Sch54,Gol66} because the operation of invariant geometric
addition of infinitesimal coordinate segments is noncommutative. So,
such spacetimes seem {\it physically inadmissible} as this result is in
direct conflict with the operational and epistemological basis of our
difference physics \cite{Ed62}. Besides, such spacetime {\it cannot be approximated locally}
by a flat, Minkowskian spacetime already on classical level.

\item {\it Torsion is topologically trivial}. This means that the
topological invariants of a real manifold $M$ and characteristic classes
of vector bundles over $M$, as defined in \cite{Trau84,Kob63,Mau93}  depend only on curvature
and can be fully determined by the curvature $_{LC} \Omega^i_{~k}$ of the
Levi--Civita connection. Roughly speaking, one can continuously deform any
metric--compatible connection (or even general linear connection) into
Levi-Civita connection {\it without changing topological invariants and
characteristic classes}. So, torsion is not relevant for topological invariants
and characteristic classes.\footnote{Some authors say that torsion which
satisfies differential field equations might be topologically
non-trivial. But this seems to be incorrect because one can still
continuously deform the connection in the case into torsionless Levi-Civita
connection {\it without changing topological invariants and
characteristic classes}. The field equations will, of course, change
during such deformation. So, it seems to us that one can say only that the
torsion which satisfies differential field equations {\it might be
physically} non-trivial. Of course, one cannot exclude that there exist
other topological properties of spacetime which can substantially depend on torsion.}
\item {\it Torsion is not relevant from the dynamical point of view either}.
Namely, one can reformulate every metric theory of gravitation with a metric-
compatible  connection $\omega^i_{~ k}$ as a "Levi-Civita theory". Torsion is
then treated as {\it a matter field}. Such reformulation preserves the
all dynamical properties of the theory. An obvious example is given by {\bf
ECSK} theory in the so--called ``combined formulation'' \cite{Hehl76}.\footnote{In this
formulation {\bf ECSK} theory is dynamically fully equivalent to the ordinary
{\bf GR} \cite{Nes77}.}

In general, one can prove \cite{Schw80} that any total Lagrangian of the
type
\begin{equation}
L_t = L_g(\vartheta^i, \omega^i_{~ k}) + L_m (\Psi, D\Psi)
\end{equation}
{\it admits an unique decomposition} into a pure geometric part ${\tilde
L}_g(\vartheta^i, _{LC} \omega^i_{~ k})$ containing no torsion plus a
generalized matter Lagrangian ${\tilde L}_m (\Psi, _{LC}D\Psi, K^i_{~ k},
_{LC}DK^i_{~ k})$ which collects the pure matter terms and all the terms
involving torsion
\begin{equation}
L_t = L_g + L_m = {\tilde L}_g + {\tilde L}_m.
\end{equation}
Here $_{LC}D$ means the exterior covariant derivative with respect to the
Levi-Civita connection $_{LC}\omega^i_{~k}$.

From the Lagrangian
\begin{equation}
L_t = {\tilde L}_g + {\tilde L}_m
\end{equation}
there follow the {\it Levi-Civita equations associated with} $L_t$.

So, torsion {\it can always be treated as a matter field}. This point of view
is preferred e.g. in \cite{Kij82,Jak88} and it is supported by transfromational
properties of torsion: torsion transforms like a matter field i.e., it
transforms as a tensor--valued form.
\item A gravitational theory with torsion {\it violates} {\bf EEP}, which has so very good
experimental evidence. It is because in a spacetime with torsion a tangent space $T_p(M)$
{\it cannot be identified with Minkowskian spacetime}, i.e., there do not exist holonomic
frames such that $g_{ik}(P) = \eta_{ik}, ~\Gamma ^i_{~kl} =0$, and, in which
geometry, in an infinitesimal vicinity of the point $P$, is Minkowskian.
$P$ is here a preselected point.
So, a gravitational theory with torsion {\it is not a covering theory for} {\bf
SRT} \cite{Roh65} and violates {\bf EEP} (Strictly
speaking, it violates {\bf LLI}). A correct relativistic theory of
gravity should be a covering theory for the both theories, {\bf SRT}
and Newton's theory of gravity. Of course, {\bf GR} satisfies this
condition.

We also lose Fermi coordinates \cite{Sch54,Mis73,Wan79,Fermi} in a Riemann-Cartan
space-time.\footnote{ Fermi coordinates realize in {\bf GR} a local
(freely-falling and non-rotating) inertial frame along a curve in which {\bf SRT} is
valid.}

Some authors \cite{Hey75,Hehl80,Blag02} formulate {\bf EEP} in a weaker form than
the constructive Will's formulation, which we have adopted in this paper.
Namely, in their formulation this Principle reads: there exists
(anholonomic for a connection with torsion) {\it normal frame} $\{\vartheta^i\}$
such that in a preselected point $P$ one has
\begin{equation}
\Gamma^i_{~kl}(P) =0,~~g_{ik}(P) = \eta_{ik}.
\end{equation}

 But this {\it Equivalence Principle is a tautology} because, as it was
 showed in past \cite{Har95}, {\it every linear connection on a
 metric manifold} satisfies it.

Moreover, if the metric-compatible connection has torsion, then,
the so-called {\it transposed connection} (see, e.g.,
\cite{Tra73})
${\hat\omega}^i_{~k}(P) := \omega^i_{~k}(P) + Q^i_{~kl}(P)\vartheta^l$,
torsion $Q^i_{~kl}(P)$ and the symmetric part $\Gamma^i_{~(kl)}(P)$
of the connection $\omega^i_{~k} = \Gamma^i_{~lk}\vartheta^l$
{\it do not vanish} in $P$ even, if in $P$, $\omega^i_{~k}(P) = \Gamma^i_{~lk}(P)\vartheta^l =
0$.

In consequence, even in a {\it normal frame}, the geometry  of
tangent space $T_p(M)$ {\it is not Minkowskian} i.e., the constructive Will's formulation
of the {\bf EEP} {\it is violated}\footnote{As we have already
emphasized, Will's formulation of the {\bf EEP} has very good
experimental evidence.}

The Equivalence Principle formulated in the form (10) {\it needs
holonomic frames} in order to efectively work. Namely, in the set
of the holonomic frames {\it it chooses} a symmetric, linear connection.
Then, adding the most natural {\it metricity postulate} (or
Hamiltonian Principle for trajectories of the test particles)
{\it univocally} leads us to (pseudo)-Riemannian geometry i.e., to the
Levi-Civita connection.

\item A connection having torsion can be determined neither by its own
autoparallells (paths) nor by geodesics \cite{Sch54}. So, one cannot determine unequivocally a
connection which has torsion by observation of the test particles (which
could move along geodesics or autoparallels).
\item Study  of the Einsteinian strength of the field equations of the proposed  gravity
theories {\it favorize} the purely metric theories of gravity (obtained with the help of
Hilbert variational principle) which use Levi-Civita connection, $_{LC}\omega$, in comparison
with competitive {\it Palatini's} theories of gravity (apart from {\bf ECSK} theory) which use
metric-compatible connection admitting torsion (see, e.g., \cite{Gar03}.
Namely, the purely metric gravity theories have {\it much more smaller} strengths
(48 in four dimensions) and numbers of dynamical degrees
of freedom (16 in four dimenions) than the competitive {\it Palatini's} {\bf PGT}
(120 and 40 in four dimensions respectively).

Following Einstein, from the two competitive gravity theories this
one is better, which has smaller strength and smaller number
dynamical degrees of freedom because such theory determines gravitational
field {\it more precisely}. More precisely in the sense: it admits
a {\it smaller number} of arbitrary initial data (putting in ``by hand'')
in the Cauchy problem, i.e., it admits smaller freedom in obtaining
a solution to the field equations.
\item Reduction of the principal bundle of the linear frames $L[M_n,~GL(n;r),~\pi]$
over $M_n$ to subbundle of the (pseudo)orthonormal frames $O[M_n,~O(n;k)~\pi]$
\footnote{ For $n =4,~k=1$ one has Lorentz group $L$.} {\it leads us
univocally} to the Levi-Civita connection. Namely, we have the
Theorem \cite{SulW}.

Theorem

Let $[M_n,~g]$ be a (pseudo)Riemannian manifold of an arbitrary
signature, $k$,. Then, there exists {\it one and only one} linear
connection $\omega$ on $L[M_n,~GL(n;r),~\pi]$ with {\it null
torsion} $\Theta =D\theta =0$ which can be reduced to the group
$O(n;k)$, i.e., to the connection $\omega_R$ on the principial
bundle $[M_n,~O(n;k),~\pi]$.

Interestingly, that $\omega$, and reduced connection $\omega_R$, {\it
are exactly the Levi-Civita connection}  $_{LC}\omega$ for the
metric $g$.

So, the fibre bundle approach suggests choosing of the symmetric
and metric Levi-Civita connection for the mathematical model $M_4(g_l,~\Gamma)$
of the physical spacetime.
\end{enumerate}

Torsion leads to ambiguities:
\begin{enumerate}
\item The {\it Minimal Coupling Principle} ({\bf MCP}) differs from the {\it
Minimal Action Principle} ({\bf MAP}) in a spacetime with torsion \cite{Sab85}.

The {\bf MCP} can be formulated as follows. In {\bf SRT} field equations
obtained from the {\bf SRT} Lagrangian density $L = L(\Psi ,
\partial_i\Psi)$ we replace
$\partial_i\longrightarrow\nabla_i,~~\eta_{ik}\longrightarrow g_{ik}$
and get {\it covariant field equations} in $(M_4,g)$.

By the {\bf MAP} we mean an application of the Minimal Action Principle
(Hamiltonian Principle) to the covariant action integral $S =
\int\limits_{\Omega} L(\Psi,~ D\Psi)d^4\Omega$, where $L(\Psi,
~ D\Psi)$ is a covariant Lagrangian density obtained from the {\bf
SRT} Lagrangian density $L(\Psi,~\partial_i\Psi)$ by {\bf MCP}.

It is natural to expect that the field equations in $(M_4,g)$ obtained by
using {\bf MCP} on {\bf SRT} equations should coincide with the
Euler-Lagrange equations obtained from $L(\Psi,~ D\Psi)$ by {\bf
MAP}. {\it This holds in GR but not in the framework of the
Riemann-Cartan geometry}. So, we have there an ambiguity of the field equations.\footnote{Axial
torsion removes this ambiguity. By $(M_4,g)$ we mean here a general metric
manifold; not necessarily Riemannian.}
\item In the framework of the {\bf ECSK} theory of gravity we have four
energy-momentum tensors for matter: Hilbert, canonical, combined, formal
\cite{Hehl76}. Which one is more important?
\item Let us consider now normal coordinates {\bf NC(P)} \cite{Sch54,Pet66,Bre98,Nest99} which
are so very important in {\bf GR} (see, eg., \cite{Pet66,Bre98,Nest99,Mul97}). In the framework of the Riemann-Cartan geometry we have
two {\bf NC(P)}: normal coordinates for the Levi-Civita part of the
Riemann-Cartan connection ${\bf NC(_{LC}\omega,~P)}$ and normal
coordinates for the symmetric part of the full connection ${\bf
NC(_s\omega,~P)}$ \cite{Gar78}. Which one has a greater physical meaning?

The above ambiguity of the normal coordinates\footnote{Axial torsion
removes this ambiguity.} leads us to ambiguities in
superenergy and supermomentum tensors \cite{Gar78}. Moreover, the obtained
expressions are too complicated for practical use. In fact, we lose
here a possibility of effective use of the normal coordinates
which give a very powerful tool in {\bf GR} to extract physical
content hidden in various non-covariant expressions.

Perhaps by use {\it normal frames} defined in \cite{Har95, Nes10} instead of normal
coordinates one could avoid these ambiguities and connected
problems. This conjecture will be studied in future.

\item In the framework of Riemann-Cartan geometry \cite{Sch54} there holds
\begin{equation}
R_{(ik)lm} = R_{ik(lm)} = 0,
\end{equation}
but
\begin{equation}
R_{iklm}\not= R_{lmik}.
\end{equation}
The last asymmetry leads to an ambiguity in construction of the
so--called ``Maxwellian superenergy tensor'' for the field $R_{iklm}$
\cite{Gar78}. This tensor is uniquely constructed in {\bf GR} owing to the
symmetry $R_{iklm} = R_{lmik}$ and it is proportional to the
Bel-Robinson tensor \cite{Gar78}. In the framework
of the Riemann-Cartan geometry the obtained result depends on which antisymmetric pair of the
$R_{iklm}$, the first or second, is used in the construction.
\item In a Riemann-Cartan spacetime we have geodesics and
autoparallells (paths). Hamiltonian Principle demands geodesics as
trajectories for the test particles \cite{Roh65}. Then, what about the physical
meaning of the autoparallells? \footnote{Axial torsion removes this
problem. One can also easily prove in the framework of the {\bf ECSK} theory
that spinless test particles move along geodesics.}
\item In a spacetime with torsion we have in fact three kinds of parallel displacement
defined by
\begin{equation}
dv^k = (-)\Gamma^k_{ij}v^j dx^i,
\end{equation}
\begin{equation}
dv^k = (-)\Gamma^k_{ij}v^i dx^j,
\end{equation}
and
\begin{equation}
dv^k = (-) \Gamma^k_{~(ij)} v^idx^j,
\end{equation}
and three different curvatures.
These results  follow from that three kinds of covariant (and absolute) differentials
\begin{equation}
\nabla^{(L)}_i{} v^k = \partial_i v^k + \Gamma^k_{il}v^l,
\end{equation}
\begin{equation}
\nabla^{(R)}_i {}v^k = \partial_i v^k + \Gamma^k_{li}v^l,
\end{equation}
\begin{equation}
\nabla^{(s)}_{i}{}v^k = \partial_i v^k + \Gamma^k_{~(li)} v^l.
\end{equation}

Authors usually use only one of the two first possibilities. What
about the others?

In a torsionless spacetime the above three possibilities coincide.

The ambiguities (13),(14)---(16),(17) arise from the two possibilities expanding of
the local connection forms ${\tilde\omega}^i_{~k}$ on the base
space $M_n$ in coordinate frames:
\begin{equation}
{\tilde\omega}^i_{~k} = \Gamma^i_{~kl}dx^l,
~~{\tilde\omega}^i_{~k} = \Gamma^i_{~lk}dx^l.
\end{equation}
\end{enumerate}

In practice, one must consequently use one of the two above
possibilities (or conventions) in order to avoid mistakes.
\subsection{Symmetry of the energy--momentum tensor of matter.}
In Special Relativity ({\bf SRT}) the  correct energy--momentum tensor for
matter (electromagnetic field,continuous medium, dust, elastic body, solids) {\it must be
symmetric} \cite{Mis73,Sch85}.

One can always get such a tensor starting from the {\it canonical
pair} $_c T^{ik}, _c  S^{ikl}= (-) _c S^{kil}$, where $_c T^{ik}\not= _c
T^{ki}$ is the canonical energy-momentum tensor and $_c S^{ikl}$ --- the
canonical spintensor. These two canonical tensors are connected by the
equations
\begin{equation}
\partial_k {_c T^{ik}} = 0,~ ~ _c T^{ik} - _c T^{ki} = \partial_l S^{ikl}.
\end{equation}
By use of the {\it Belinfante symmetrization procedure}
\cite{Hehl76,Sab85,Bel39,Kop84} one can get the most simple new pair
\begin{equation}
_s T^{ik} = _c T^{ik} - {1\over 2}\partial_j\bigl(_c S^{ikj} - _c S^{ijk} + _c
S^{jki}\bigr),
\end{equation}
\begin{equation}
S^{ijk} = _c S^{ijk} - A^{jki} + A^{ikj} = 0.
\end{equation}
Here
\begin{equation}
A^{ikj} = {1\over 2}\bigl(_c S^{ikj} - _c S^{ijk} + _c S^{jki}\bigr).
\end{equation}
The obtained new "pair" $(_sT^{ik}, ~ 0)$ is {\it the most simple and the
most symmetric}. Note that the symmetric tensor $_sT^{ik} = _s T^{ki}$
gives complete description of matter because the spin density tensor $_c S^{ijk}$
is {\it entirely absorbed} into $_s T^{ik}$ by the symmetrization procedure.

Note also that the symmetric tensor $_s T^{ik}$ has 10 independent
components and this number is exactly the same as the number of
integral conserved quantities in an asymptotically flat closed
system.

It is interesting that one can easily generalize the above symmetrization
procedure onto a general metric manifold $(M_4, g)$ \cite{Gren83,Hehl76} by using the Levi-Civita connection associated with the
metric $g$. The generalized symmetrization procedure has the same form as above
with the replacement $\eta_{ik} \longrightarrow g_{ik},~~\partial_i
\longrightarrow _{LC} \nabla_i$.

So, one can always get on a metric manifold $(M_4,g)$ a symmetric
energy--momentum tensor $_sT^{ik} = _sT^{ki}$ for matter (then, of course, corresponding
$S^{ikj} = 0$). Observe that the symmetric tensor $_sT^{ik}$, like as in {\bf SRT},
consists of the canonical tensors $_cT^{ik}$ and $_cS^{ikl}$.

The symmetric energy--momentum tensor for matter {\it is unique}, i.e., it is uniquely
determined by the matter equations of motion and reasonable boundary conditions \cite{Foc69}.
This fact is essential for the uniqueness of the gravitational field equations. Moreover, the
symetric energy--momentum tensor is covariantly conserved (a canonical energy-momentum tensor
is not conserved).

L. Rosenfeld has proved \cite{Ros40} that
\begin{equation}
_s T^{ik} = {\delta L_m\over\delta g_{ik}},
\end{equation}
where $L_m = L_m(\Psi,~ _{LC}D\Psi)$ is a covariant Lagrangian density for
matter. The tensor $_s T^{ik}$ given by (24) is the source in the
Einstein equations
\begin{equation}
G_{ik} = \chi _s T_{ik},
\end{equation}
where $\chi = {8\pi G\over c^4}$.

Note that these equations {\it geometrize both the canonical quantities}
$_c T^{ik}$ and $_c S^{ikl} = (-)_c S^{kil}$ {\it in some equivalent
way} because  the tensor $_s T^{ik}$ is built from these two canonical
tensors.

So, it is the most natural and most simple to postulate that, in general,
the correct energy--momentum tensor for matter is the symmetric tensor $_sT^{ik}$.
This leads us to a purely metric torsion-free theory of gravity with
the field equations
\begin{equation}
{\delta L_g\over\delta g_{ik}} = {\delta L_m\over\delta g_{ik}}.
\end{equation}
Then, if we take into account the {\it dynamical universality} of the Einstein
equations  \cite{Jak88,Mag87,Bor94}, we will end up with General Relativity (possibly with
$\Lambda \not= 0$) which will have a sophisticated, symmetric energy-momentum
tensor as a source.
\subsection{Some remarks on the ``teleparallel equivalent of general relativity''}

After presenting the preliminary draft of the old our lectures in
arXiv \cite{Jangar}, we have got critical remarks from some persons which are working on
the so-called {\it teleparallel equivalent of general relativity} ({\bf TEGR}) in the framework
 of the {\it Weitzenb\"ock or teleparallel geometry} \cite{Sch54,Gol66,And00}.
Our reply was the following \footnote{This reply was considerably
extended and updated in the paper \cite{Gar10}.}.
The {\it Weitzenb\"ock or teleparallel connection and geometric
structure} on spacetime is determined by a tetrad (or other anholonomic
frame) field $h^{(a)}_{~~~b}(x)$ and  {\it can always} be introduced
{\it independently} of the geometric structure of the
spacetime.  Here $(a),(b),...$ are tetrad (= anholonomic) indices and
$a,b,c,...$ mean holonomic (= world) indices.

The fundamental formulas of the teleparallel geometry read
\begin{equation}
g_{ik} := \eta_{(a)(b)} h^{(a)}_{~~~i} h^{(b)}_{~~~k},
\end{equation}
\begin{equation}
\Gamma^i_{~kl} := h_{(a)}^{~~~i}{} \partial_k h^{(a)}_{~~~l},
\end{equation}
\begin{equation}
\nabla_i h^{(a)}_{~~~k} = 0,
\end{equation}
\begin{equation}
\Gamma^i_{~kl} = _{LC} \Gamma^i_{~kl} + K^i_{~kl},
\end{equation}
\begin{equation}
K^i_{~kl} := 1/2\bigl({T_k^{~i}}{}_l + {T_l^{~i}}{}_k - T^i_{~kl}\bigr),
\end{equation}
\begin{equation}
T^i_{~kl} := \Gamma^i_{~lk} - \Gamma^i_{~kl},
\end{equation}
and
\begin{equation}
R^i_{~klm} = _{LC} R^i_{~klm} + Q^i_{~klm} \equiv 0,
\end{equation}
where $Q^i_{~klm}$ is a tensor written in terms of the {\it contortion}
$K^i_{~kl}$ and its covariant derivatives with respect to the
Levi-Civita connection $_{LC} \Gamma^i_{~kl}$ of the metric $g_{ik}$.

Here $\eta_{(a)(b)}$ means the {\it interior metric} (usually
Minkowskian) of a tangent space and the duals $h_{(a)}^{~~~i}$ are defined by
\begin{equation}
h_{(a)}^{~~~i}{} h^{(a)}_{~~~k} = \delta^i_k.
\end{equation}

Those authors which work on {\bf TEGR}, by use the formulas (27), (30), and (33)
of the teleparallel geometry {\it rephrase}, step-by-step, the all
formalism of {\bf GR} in terms of the Weitzenb\"ock connection
$\Gamma^i_{~kl}$ and its {\it torsion} $T^i_{~kl}$. Then, they call this
formal reformulation  of {\bf GR} in terms of the Weitzenb\"ock geometry
{\it the teleparallel equivalent of general relativity} ({\bf TEGR})
(What kind of ``equivalence''?).

One can read in the papers \cite{And00} the following conclusion:
"Gravitational interaction, thus, can be described alternatively in
terms of curvature, as is usually done in {\bf GR}, or in terms of
torsion, in which case we have the so--called {\it teleparallel
gravity}. Whether gravitation requires a curved or torsional spacetime,
therefore, turns out to be a matter of convention".

From the point of view of the {\bf TEGR}, therefore, teleparallel torsion has
fundamental physical meaning and it has been already  detected.

We cannot agree with such statements. In our opinion, the "teleparallel
equivalent of {\bf GR}" is only {\it formal} and geometrically {\it trivial}
rephrase of {\bf GR} in terms of the Weitzenb\"ock geometry. Such
rephrase is, of course, {\it always possible} not only with {\bf GR} but
also with any other purely metric theory of gravity (see eg. \cite{Schim02})
but it {\it has no profound physical motivation}. It is because, as one can easily show,
the teleparallel torsion is entirely expressed in terms of the Van Danzig and Schouten
aholonomity object $\Omega^{(a)}_{~~~(b)(c)}$ (see eg. \cite{Sch54,Gol66}). So, the
torsion $T^i_{~kl}$ of a teleparallel connection {\it describes only
anholonomity} of the used field of aholonomic frames $h^{(a)}_{~~~i}(x)$;
{\it not real geometry of the spacetime}. \footnote{Unless one can
physically distinguish a tetrad field (or other anholonomic field of frames) and give it a
fundamental geometrical and physical meaning. But we think that this
could introduce a cristal--like structure on spacetime and, therefore, it
would contradict local Lorentz invariance.}.
Contrary,  Levi-Civita part of a Weitzenb\"ock connection can have (and has)
geometrical (and physical) meaning.

Resuming, it seems to us that {\bf TEGR} is rather  a {\it mathematical
curiosity} which gives, by no means, {\it anything better} than ordinary
{\bf GR} gives and one can doubt into its physical meaning.

Precise experimental confirmation of the {\bf EEP} proved non-zero
curvature of physical spacetime \cite{Mis73,Schi67} and supported ordinary {\bf GR}. We
think that this fact excludes a physical motivation  for {\it rephrasing}
{\bf GR} into {\bf TEGR}.

One remark more is in order concerning {\bf TEGR}: {\bf TEGR}
resulted in $f(T)$ theories where $T$ means the Lagrangian density
\cite{2011} of the {\bf TEGR}. In analogy to $f(R)$ extension of
the Hilbert action of {\bf GR}, the $f(T)$ theories  are
generalization of the action of {\bf TEGR}.

It seems that the only one positive property of these theories is
the fact that they have 2-nd order field equations.

\section{Concluding remarks}
The {\bf GR} model of the space-time has very good
experimental confirmation in a weak-field approximation (Solar System)
and in the strong fields (binary pulsars). On the other hand, torsion has no
experimental evidence (at least in vacuum) and it is not needed in a
theory of gravity. Moreover, the introduction  of torsion into the
geometric structure of space-time leads to many problems (apart
from calculational, of course). Most of these problems are removed if
only axial torsion $A_i = {1\over 6}\eta_{iabc}Q^{abc}, ~Q^{[abc]} =
Q^{abc}$ exists. So, it would be reasonable to confine themselves to the axial torsion only
(If one still want to keep on torsion). This is also supported by the
important fact that the {\it matter fields} (= Dirac's particles) are
coupled only to the axial part of torsion in the Riemann-Cartan space-time.

However, if we confine to the axial torsion, then (if we remember the dynamical triviality
of torsion and the dynamical universality of the Einstein equations) we
effectively will end up with {\bf GR} + additional matter fields. In the
most important case of the {\bf ECSK} theory we will end up with {\bf
GR} + an aditional pseudovector field $A_i$ (or with
an additional pseudoscalar field $\varphi$ if the field $A_i$ is
potential, i.e., if $A_i = \partial_i\varphi$) \cite{Sab85}. But {\bf GR} with
an additional dynamical pseudovector field $A_i$ yields local
gravitational physics which may have both location and
velocity-dependent effects \cite{Clif93} unobserved up to now. Besides, {\bf GR} with
an additional pseudoscalar field has a defect because there exist two
distinguished frames, {\it the Einstein frame} and {\it the Jordan frame}, which are
not equivalent physically \cite{Mac01}.

Additionally, we would like to emphasize that there exist very strong experimental
constraints on the components of the axial torsion: $<10^{(-15)} m^{(-)1}$ \cite{Shap}.

So, we will finish with the conclusion that the geometric model of the
space-time given by ordinary {\bf GR} and ``wonderful'' Levi-Civita
connection {\it seems to be the most satisfactory}.

Interestingly that this model has a very strong support
from the field-theoretic approach to gravity (see e.g., \cite{Str00}).

It seems to us that the torsion was introduced into a theory of
gravity in order to get some link between theory of gravity and quantum
fundamental particles theory (It is commonly known that the role of the
curvature in an atomic and smaller scale is neglegible). But these
trials {\it were not successful} (see, e.g., \cite{Shap}). It also seems that what we
really need nowadays is a {\it quantum model} of the Riemannian geometry and a
{\it quantum gravity} which is based on this model. The recent papers given by Ashtekhar
\cite{Ash99,Iri88,Lew04,Rov20}
and co--workers on this problem seems to be very promissing.
\vspace{.3cm}
\begin{center}
{\bf Acknowledgments}
\end{center}
\vspace{.2cm}

The author would like to thank Prof. J. \L awrynowicz for possibility to deliver
this lecture at the {\bf Hypercomplex Seminar 2011} in B\c edlewo,
Poland, and the Mathematical Institute of the University of
Szczecin for financial support (grant 503-4000-230351).

\newpage
\begin{center}
{\bf Czy torsja jest potrzebna w teorii grawitacji? Nowe
Spojrzenie}
\end{center}
\vspace{.3cm}
\begin{center}
Janusz Garecki\\
Instytut Matematyki Uniwersytetu Szczeci\'nskiego\\
Streszczenie
\end{center}
\vspace{.3cm}

W pracy pokazano, \.ze wprowadzenie skr\c ecenia do modelu
matematycznego fizycznej czasoprzestrzeni nie jest {\it ani konieczne,
ani wskazane}.

Model matematyczny, kt\'ory daje og\'olna teoria wzgl\c edno\'sci
{\it jest wystarczaj\c acy} dla wszelkich potrzeb fizyki i, jak dot\c
ad, jest {\it bardzo dobrze} potwierdzony przez eksperymenty.
\end{document}